\documentclass[aps,pra,preprint,groupedaddress]{revtex4}
\usepackage{graphicx}

\begin{document}

\title{Dynamical Instabilities in a two-component Bose condensate in a 1d optical lattice}

\author{Samantha Hooley and Keith A Benedict}
\affiliation{Condensed Matter Theory, School of Physics and Astronomy,\\ University of Nottingham, University Park, \\NOTTINGHAM, NG7 2RD, UK}

\date{\today}

\begin{abstract}
In this paper we carry out a stability analysis of the Bloch states of a two-component Bose-Einstein condensate confined to a 1d optical lattice. We consider two concrete systems: a mixture of two hyperfine states of Rubidium-87 and a mixture of Sodium-23 and Rubidium-87. The former is seen to exhibit similar phenomena to a single component condensate while the latter also suffers an instability to phase separation at small Bloch wave vectors. It is shown that sufficiently deep optical lattices can remove this latter instability, potentially allowing imiscible cold atoms species to be held in intimate contact and transported within an experimental system.
\end{abstract}


\maketitle
\section{Introduction}

Bloch states of weakly interacting Bose condensates have been the subject of
recent experimental and theoretical study (see \cite{MorschReview} for a review). While these states are, in many
respects, similar to those of electrons in solids, they differ in one crucial
respect - they are not necessarily stable \cite{WuNiu, Machholm}. Isolated, weakly interacting atomic
condensates are well described by mean field theory in the form of the
time-dependent Gross-Pitaevskii (GP) equation. In the presence of an optical
lattice this has stationary Bloch solutions of the form%
\begin{equation}
\Psi\left(  x,t\right)  =e^{-i\mu t/\hbar}e^{ikx}f\left(  x\right)
\end{equation}
where $f\left(  x\right)  $ has the periodicity of the lattice and
$\mu$ is the chemical potential for the atoms. Many of the phenomena
associated with these Bloch states are familiar from the theory of electrons
in solids (band structure etc.). Indeed, cold atoms systems readily exhibit
phenomena which are hard to observe in electronic systems such as Bloch
oscillations. However, unlike electronic states in a crystalline lattice, the
Bloch solutions of the GP\ equation are not necessarily stable. As shown by Wu
and Niu and Machholm, Pethick and Smith\cite{WuNiu,Machholm}, there are two types of instability. Firstly
there is an energetic instability associated with Bloch states which are not
local minima of the mean field energy, secondly there is a dynamical
instability associated with the exponential growth of small perturbations
around the Bloch state. The former instability should not be visible in
systems which are perfectly described by the GP equation, which conserves the
mean field energy, but will have an effect once dissipative processes become
active. The dynamical instability is always significant unless the time for
which the system is described by the unstable Bloch state is much shorter than
the growth time for the most unstable mode.

In this paper we consider the dynamical stability of a system consisting of
two distinct atomic species. The laser generating the standing wave optical
lattice is supposed to have been chosen so that it is blue detuned from the
nearest resonance of one species but red-detuned from the nearest resonance of
the other. This means that the two species will see opposite potentials as one
is attracted to the nodes of the standing wave while the other is attracted to
the antinodes. We also suppose that the system is strongly confined by a
cylindrically symmetric magnetic trap to ensure one dimensional behaviour.

In section 2 we review the mean field theory for such a system and the
2-component GP\ equation. In section 3 we consider the Bloch states of the
2-component condensate. In section 4 we review the linear stability analysis
of the mean field theory. In section 5 we consider two specific
systems: firstly we consider the two species to be 2 hyperfine states of
Rubidium; next we consider a mixture of Rubidium and Sodium atoms. We will see
that these two examples exhibit rather different behaviour, the former is
qualitatively similar to a single component system while the later exhibits
quite different behaviour. In section 6 we will discuss the results obtained.

\section{Mean Field Theory for 2-Component System}
The mean field theory for a two component system is well described in the book
by Pitaevskii and Stringari\cite{PitaevskiiBook}. The main result is that the two component
condensate wave function, $\Psi_{j}\left(  x,t\right)  $ satisfies the
coupled GP\ equations%
\begin{eqnarray}
i\hbar\frac{\partial\Psi_{1}}{\partial t} &  = &-\frac{\hbar^{2}}{2m_{1}}%
\frac{\partial^{2}\Psi_{1}}{\partial x^{2}}+V_{1}\left(  x\right)  \Psi
_{1}+g_{11}\left|  \Psi_{1}\right|  ^{2}\Psi_{1}+g_{12}\left|  \Psi
_{2}\right|  ^{2}\Psi_{1}\\
i\hbar\frac{\partial\Psi_{2}}{\partial t} &  = & -\frac{\hbar^{2}}{2m_{2}}%
\frac{\partial^{2}\Psi_{2}}{\partial x^{2}}+V_{2}\left(  x\right)  \Psi
_{2}+g_{22}\left|  \Psi_{2}\right|  ^{2}\Psi_{2}+g_{12}\left|  \Psi
_{1}\right|  ^{2}\Psi_{2}%
\end{eqnarray}
where $m_{j}$ is the mass of an atom of species $j$ and $V_{j}$ is the optical
lattice potential seen by species $j$. The nonlinear terms arise from the
treatment of the atomic collisions within mean field theory. For the low
temperatures relevant to ultra-cold atom experiments it is sufficient to treat
the atoms as point scatterers with strengths given in terms of the relevant
s-wave scattering length within the lowest Born approximation. This gives%
\begin{equation}
g_{ii}=\gamma_{ii}\frac{4\pi\hbar^{2}a_{i}}{m_{i}}%
\end{equation}
for 2-body processes in which like atoms scatter and%
\begin{equation}
g_{12}=g_{21}=\gamma_{12}\frac{2\pi\hbar^{2}a_{12}}{m_{12}}%
\end{equation}
for 2-body processes in which unlike atoms scatter. The $\gamma_{ij}$'s are form-factors accounting for the finite extent of the wavefunctions transverse to the optical lattice. For a harmonic radial trap of the form $U\left(r\right)=C^2r^2/2$ we have
\begin{equation}
\gamma_{ij}=\frac{1}{\pi \left(l_i^2+l_j^2\right)}
\end{equation}
where the radial oscillator lengths for the two species are $l_i^2=\hbar/C\sqrt{m_i}$. The expression for $g_{12}$ involves the reduced mass $m_{12}=\left(  m_{1}^{-1}+m_{2}^{-1}\right)  ^{-1}$
relevant for the ``unlike'' 2-body problem.

Stationary solutions of these equations of the form $\Psi_{j}\left(
x,t\right)  =e^{-i\mu_j t/\hbar}\psi_{j}\left(  x\right)  $ satisfy the
time-independent GP\ equations%
\begin{eqnarray}
-\frac{\hbar^{2}}{2m_{1}}\frac{d^{2}\psi_{1}}{dx^{2}}+V_{1}\left(  x\right)
\psi_{1}+g_{1}\left|  \psi_{1}\right|  ^{2}\psi_{1}+g_{12}\left|  \psi
_{2}\right|  ^{2}\psi_{1}  &  =\mu_1\psi_{1}\\
-\frac{\hbar^{2}}{2m_{2}}\frac{d^{2}\psi_{2}}{dx^{2}}+V_{2}\left(  x\right)
\psi_{2}+g_{2}\left|  \psi_{2}\right|  ^{2}\psi_{2}+g_{12}\left|  \psi
_{1}\right|  ^{2}\psi_{2}  &  =\mu_2\psi_{2}%
\end{eqnarray}
in which the chemical potentials of the components appear as the solution to a
non-linear eigenvalue problem.

The time-independent GP equation can be obtained variationally from the mean
field energy functional%
\begin{equation}
\mathcal{E}=\int dx\left\{  \sum_{j}\left(  \frac{\hbar^{2}}{2m_{j}}\left|
\frac{d\psi_{j}}{dx}\right|  ^{2}+V_{j}\left(  x\right)  \left|  \psi
_{j}\right|  ^{2}\right)  +\frac{1}{2}\sum_{ij}g_{ij}\left|  \psi_{i}\right|
^{2}\left|  \psi_{j}\right|  ^{2}\right\}
\end{equation}
via%
\begin{equation}
\frac{\delta}{\delta\psi_{j}^{\ast}\left(  x\right)  }\left(  \mathcal{E}%
-\sum_{j}\mu_j\int dx\left|  \psi_{j}\right|  ^{2}\right)
\end{equation}
where, as usual, the chemical potentials arise as a Lagrange multipliers
enforcing the normalization conditions%
\begin{equation}
\int dx\left|  \psi_{j}\right|  ^{2}=N_{j}%
\end{equation}
for a system with $N_{j}$ atoms of species $j$.

The chemical potential of a stationary state is related to the mean field
energy via the relation%
\begin{equation}
\mathcal{E}=\left(  \mu_1N_{1}+\mu_2N_{2}\right)  -\sum_{ij}\frac{1}{2}g_{ij}\int
dx\left|  \psi_{i}\right|  ^{2}\left|  \psi_{j}\right|  ^{2}\qquad.
\end{equation}

We consider a system in an infinite optical lattice with%
\begin{equation}
V_{j}\left(  x\right)  =w_{j}\cos\left(  \kappa x\right)
\end{equation}
where $\kappa=2\pi/d$ where $d=\lambda/2$ is the period of the lattice (half
of the wavelength of the laser generating the standing wave).

\section{Bloch States}

Next we seek Bloch states of the form%
\begin{equation}
\phi_{j}^{\left(  k\right)  }\left(  x\right)  =e^{ikx}f_{j}\left(  x\right)
\end{equation}
where%
\begin{equation}
f_{j}\left(  x+d\right)  =f_{j}\left(  x\right)  \qquad.
\end{equation}
We employ the same basic method as Machholm and Smith \cite{Machholm}, adapted to the case of
two-components. If $n_{j}$ is the number of atoms of species $j$ per unit
length then the normalization condition on $f_{j}(x)$ becomes%
\begin{equation}
\frac{1}{d}\int_{-d/2}^{d/2}dx\left|  f_{j}(x)\right|  ^{2}=n_{j}\qquad.
\end{equation}
We also define the energy per unit length%
\begin{eqnarray}
E & = &\frac{1}{d}\int_{-d/2}^{d/2}dx\left\{  \sum_{j}\left(  \frac{\hbar^{2}%
}{2m_{j}}\left|  \left(  \frac{d}{dx}+ik\right)  f_{j}\left(  x\right)
\right|  ^{2}+w_{j}\cos\left(  \kappa x\right)  \left|  f_{j}\left(  x\right)
\right|  ^{2}\right)\right.\\
& + & \left.\frac{1}{2}\sum_{ij}g_{ij}\left|  f_{i}\left(
x\right)  \right|  ^{2}\left|  f_{j}\left(  x\right)  \right|  ^{2}\right\}
\qquad.
\end{eqnarray}
The periodicity of $f_{j}\left(  x\right)  $ allows us to write%
\begin{equation}
f_{j}\left(  x\right)  =\sum_{s}f_{j,s}e^{is\kappa x}%
\end{equation}
so that%
\begin{equation}
\sum_{s}\left|  f_{j,s}\right|  ^{2}=n_{j}\label{constraint}
\end{equation}
and
\begin{eqnarray}
E & = & \sum_{j}\left(  \frac{\hbar^{2}}{2m_{j}}\sum_{s}\left|  f_{j,s}\right|
^{2}\left(  m\kappa+k\right)  ^{2}+\frac{w_{j}}{2}\sum_{s}\left(
f_{j,s+1}^{\ast}+f_{j,s-1}^{\ast}\right)  f_{j,s}\right)  +\\
& + &\frac{1}{2}%
\sum_{i,j}g_{ij}\sum_{s}\left(  \sum_{l}f_{i,l+s}^{\ast}f_{i,l}\right)
\left(  \sum_{l}f_{j,l-s}^{\ast}f_{j,l}\right)  \qquad.
\end{eqnarray}
As found by Machholm and Smith for the single component case, we can restrict
our attention to real values of the $f$ parameters and truncate summations
over the Fourier label to values less than a cut-off $S$. We then minimize $E$
as a function of the $2\times\left(2S+1\right)$ $f$ parameters subject to the constraint (\ref{constraint}) to find the Bloch wave functions for each component for a grid of points in the range
$-\kappa/2<k<\kappa/2$ (i.e. within the first Brillouin zone of the optical lattice).

\section{Stability Analysis}
Given the Bloch state for a particular value of $k$, we can ask whether it is
stable with respect to small fluctuations. We assume an initially small, generic
fluctuation, so that the condensate wave function has the form%
\begin{equation}
\Psi_{j}\left(  x,t\right)  =e^{-i\mu_j t/\hbar}\left(  \phi_{j}\left(
x\right)  +\delta\Psi_{j}\left(  x,t\right)  \right)
\end{equation}
which we substitute into the time-dependent GP\ equation. Dropping terms
non-linear in the fluctuations then gives
\begin{eqnarray}
i\hbar\frac{\partial\delta\Psi_{1}  }{\partial t}  &
=&\left(  -\frac{\hbar^{2}}{2m_{1}}\frac{\partial^{2}}{\partial x^{2}}%
+V_{1}\left(  x\right)  -\mu_{1}+2g_{11}\left|  \phi_{1}\left(  x\right)
\right|  ^{2}+g_{12}\left|  \phi_{2}\left(  x\right)  \right|  ^{2}\right)
\delta\Psi_{1} \\
&  +& g_{11}\left(  \phi_{1}\left(  x\right)  \right)  ^{2}\delta\Psi_{1}^{\ast
}  +g_{12}\phi_{2}^{\ast}\left(  x\right)  \phi_{1}\left(
x\right)  \delta\Psi_{2}  +g_{12}\phi_{2}\left(  x\right)
\phi_{1}\left(  x\right)  \delta\Psi_{2}^{\ast} \\
i\hbar\frac{\partial\delta\Psi_{2}  }{\partial t}  &
=&\left(  -\frac{\hbar^{2}}{2m_{2}}\frac{\partial^{2}}{\partial x^{2}}%
+V_{2}\left(  x\right)  -\mu_{2}+2g_{22}\left|  \phi_{2}\left(  x\right)
\right|  ^{2}+g_{12}\left|  \phi_{1}\left(  x\right)  \right|  ^{2}\right)
\delta\Psi_{2} \\
&  +&g_{22}\left(  \phi_{2}\left(  x\right)  \right)  ^{2}\delta\Psi_{2}^{\ast
}  +g_{12}\phi_{1}^{\ast}\left(  x\right)  \phi_{2}\left(
x\right)  \delta\Psi_{1}  +g_{12}\phi_{1}\left(  x\right)
\phi_{2}\left(  x\right)  \delta\Psi_{1}^{\ast}\qquad .
\end{eqnarray}
As expected, the $\delta\Psi$'s are coupled to their complex conjugates and we
must decouple them using a classical version of the Bogoliubov transformation,
as developed by Pitaevskii\cite{Pitaveskii61}. We set%
\begin{equation}
\delta\Psi_{j}\left(  x,t\right)  =e^{ikx}\left(  e^{i\left(  qx-\omega
t\right)  }u_{j}\left(  x\right)  +e^{-i\left(  qx-\omega t\right)  }%
v_{j}^{\ast}\left(  x\right)  \right)
\end{equation}
where $u_{j}\left(  x\right)  $ and $v_{j}\left(  x\right)  $ all have the
periodicity of the lattice. Substituting into the time dependent equation
gives, after some manipulation, an eigenvalue problem for the $u(x)$ and $v(x)$ functions of the form
\begin{equation}
\mathcal{M}_{k,q}\left(
\begin{array}
[c]{c}%
u_{1}\left(  x\right) \\
v_{1}\left(  x\right) \\
u_{2}\left(  x\right) \\
v_{2}\left(  x\right)
\end{array}
\right)  =\omega\left(
\begin{array}
[c]{c}%
u_{1}\left(  x\right) \\
v_{1}\left(  x\right) \\
u_{2}\left(  x\right) \\
v_{2}\left(  x\right)
\end{array}
\right)
\end{equation}
where%
\begin{equation}
\mathcal{M}_{k,q}   = \left(
\begin{array}
[c]{cccc}%
\mathcal{L}_{1}^{+} & g_{11}\left(  f_{1}\left(  x\right)  \right)  ^{2} &
g_{12}f_{2}\left(  x\right)  f_{1}\left(  x\right)  & g_{12}f_{2}\left(
x\right)  f_{1}\left(  x\right) \\
-g_{11}\left(  f_{j}\left(  x\right)  \right)  ^{2} & -\mathcal{L}_{1}^{-} &
-g_{12}f_{2}\left(  x\right)  f_{1}\left(  x\right)  & -g_{12}f_{2}\left(
x\right)  f_{1}\left(  x\right) \\
g_{12}f_{1}\left(  x\right)  f_{2}\left(  x\right)  & g_{12}f_{1}\left(
x\right)  f_{2}\left(  x\right)  & \mathcal{L}_{2}^{+} & g_{22}\left(
f_{2}\left(  x\right)  \right)  ^{2}\\
-g_{12}f_{1}\left(  x\right)  f_{2}\left(  x\right)  & -g_{12}f_{1}\left(
x\right)  f_{2}\left(  x\right)  & -g_{22}\left(  f_{2}\left(  x\right)
\right)  ^{2} & -\mathcal{L}_{2}^{-}%
\end{array}
\right) 
\end{equation}
and%
\begin{eqnarray}
\mathcal{L}_{1}^{\pm}  &  = & -\frac{\hbar^{2}}{2m_{1}}\left(  \frac{\partial
}{\partial x}+i\left(  q\pm k\right)  \right)  ^{2}+V_{1}\left(  x\right)
-\mu_{1}+2g_{11}\left|  f_{1}\left(  x\right)  \right|  ^{2}+g_{12}\left|
f_{2}\left(  x\right)  \right|  ^{2}\\
\mathcal{L}_{2}^{\pm}  &  = & -\frac{\hbar^{2}}{2m_{2}}\left(  \frac{\partial
}{\partial x}+i\left(  q\pm k\right)  \right)  ^{2}+V_{2}\left(  x\right)
-\mu_{2}+2g_{22}\left|  f_{2}\left(  x\right)  \right|  ^{2}+g_{12}\left|
f_{1}\left(  x\right)  \right|  ^{2}\qquad.
\end{eqnarray}
Substitution of the truncated Fourier expansions%
\begin{eqnarray}
u_{j}\left(  x\right)   &  = & \sum_{s=-\nu}^{\nu}\alpha_{j,s}e^{is\kappa x}\\
v_{j}\left(  x\right)   &  = & \sum_{s=-\nu}^{\nu}\beta_{j,s}e^{is\kappa x}%
\end{eqnarray}
allows this to be turned into an $4\left(  2\nu+1\right)  \times
4\left(  2\nu+1\right)  $ generalized matrix eigenvalue problem. The
resultant eigenvalue problem is
non-hermitian and hence need not, in general, have real eigenvalues. As shown
in \cite{Garay} the eigenvalues either come in real pairs $\pm\omega$ or in sets of four
complex eigenvalues $\pm\omega^{\prime}\pm i\omega^{\prime\prime}$. As usual
in the Bogoliubov method these are not all independent. For real eigenvalues
only modes with positive frequency need be considered. For the case of complex
eigenvalues we may discard those with negative real parts. The presence of an
imaginary part to the eigenvalue indicates an instability of the underlying
Bloch state, since there will be a mode which, at least initially, grows exponentially in time at the rate
$\omega^{\prime\prime}$.

The eigenvalues for each $q$ value form a set corresponding to the Bloch bands of
the linearized fluctuation problem. Only the lowest two fluctuation bands exhibit non-zero imaginary
parts to the eigenvalues, so we focus attention on these. We define the {\em instability} of the system with respect to modes with wave-vector $q$ as 
\begin{equation}
\theta\left(q\right)=\sup_{j=1,2}\left\{\omega''_j(q)\right\}
\end{equation}

\section{Results of Stability Analysis}

Here we will present results for the stability of the Bloch states of
2 systems. We will compare and contrast the two systems in the next section.

Firstly, we consider a system in which the two hyperfine states $\left|
F=1,m_{F}=-1\right\rangle $ and $\left|  F=2,m_{F}=2\right\rangle $ of
Rubidium-87 are cooled in a magnetic trap to form a condensate. Such a two-component condensate was first prepared experimentally by Myatt et al\cite{Myatt}. We suppose that the atoms are confined by a strong axial magnetic trap  and a longitudinal optical lattice. We further suppose that the
laser is tuned between resonances of the two species such that $V_0=w_{1}%
=-w_{2}$ and we choose a typical value for the depth of the lattice $V_0=0.765E_{R}$ where%
\begin{equation}
E_{R}=\frac{\hbar^{2}\pi^{2}}{2md^{2}}%
\end{equation}
is the recoil energy associated with a Rb atom absorbing a photon from the
laser generating the standing wave. We also assume equal densities of the two
species, $n_{1}=n_{2}=n$ and%
\begin{eqnarray}
ng_{11}  &  = & ng_{22}=0.1E_{R}\\
ng_{12}  &  = & 0.099E_{R}%
\end{eqnarray}
so that, apart from the opposite signs of the $v_{j}$'s, the two species are
virtually identical.
\begin{figure}
\includegraphics[width=9cm]{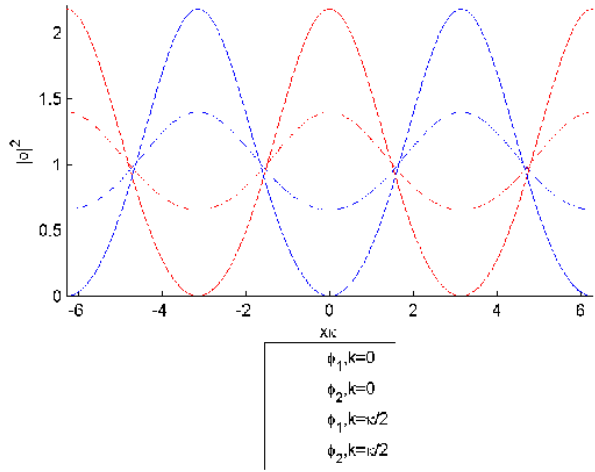}
\caption{The Bloch wavefunctions $f_j(x)$ at $k=0$ (dotted) and $k=\kappa/2$ (full) plotted against $\kappa x$\label{Fig1}}
\end{figure}
In figure \ref{Fig1} we show the form of the Bloch functions $f_{1}$
and $f_{2}$ for both $k=0$ and $k=\kappa/2$ (zone boundary). As can be seen,
the $k=0$ Bloch states are lightly modulated by the optical lattice with one
species attracted to the nodes of the optical standing wave and the other
attracted to the antinodes. The $k=\kappa/2$ Bloch states show much stronger
modulation with the wavefunctions of the two species vanishing at the nodes and antinodes respectively.
\begin{figure}
\includegraphics[width=8cm]{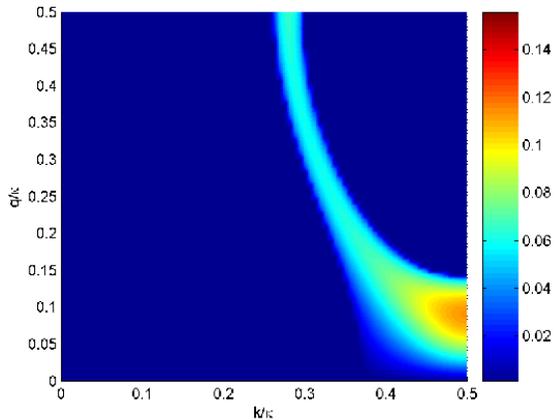}
\caption{The imaginary part of the fluctuation mode frequency, $\Im \omega^{(k)}_1(q)$ for $0<k<\kappa/2$ and $0<q<\kappa/2$ (colour online). \label{Fig7}}
\end{figure}
\begin{figure}
\includegraphics[width=8cm]{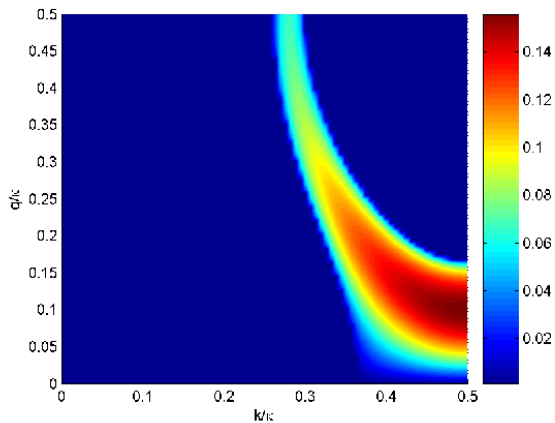}
\caption{The imaginary part of the fluctuation mode frequency, $\Im \omega^{(k)}_2(q)$ for $0<k<\kappa/2$ and $0<q<\kappa/2$ (colour online).\label{Fig8}}
\end{figure}

Figures \ref{Fig7} and \ref{Fig8} show colour maps of the imaginary part of the fluctuation frequencies, $\omega''_1(q)$ and $\omega''_2(q)$ respectively, for
the lowest two fluctuation bands as functions of the Bloch wave-vector $k$ and the
fluctuation wavevector $q$. As can be seen, both of the lowest bands exhibit
behaviour similar to that of a single component condensate. The Bloch states
are stable for $k\lesssim\kappa/4$. States with higher Bloch wave-vectors are
unstable with respect to fluctuation modes with wave-vectors around
$q=\kappa/2$:$\;$effectively a period doubling modulational instability. As
$k$ is increased, the unstable modes move to longer wavelengths with a greater
range of $q$'s being unstable. All of this behaviour is qualitatively similar
to the behaviour of a single component condensate.

Now we consider a different system: a mixture of Sodium and Rubidium atoms,
similarly confined to a 1d optical lattice tuned between resonances of the two
species. We take typical values $V_0=w_{1}=-w_{2}=0.603E_{R}$ and again assume
equal densities $n_{1}=n_{2}=n$ such that $ng_{11}=0.047E_{R}$, $ng_{22}%
=0.019E_{R}$ and $ng_{12}=0.048E_{R}$. 
\begin{figure}
\includegraphics[width=8cm]{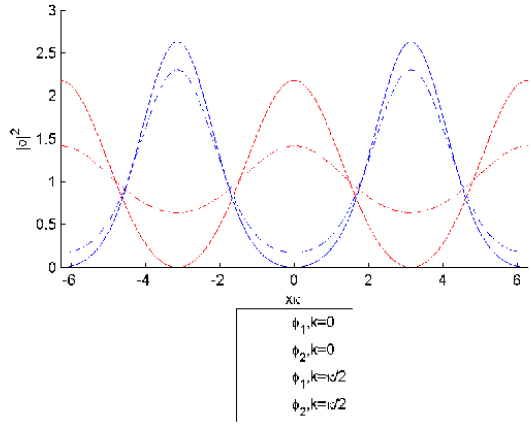}
\caption{The Bloch wavefunctions $f_1(x)$ and $f_2(x)$ for a Rb-Na mixture for $k=0$ (dotted) and $k=\kappa/2$ (full).\label{Fig10}}
\end{figure}
Figure \ref{Fig10} shows the Bloch wavefunctions
for this system at $k=0$ and $k=\kappa/2$. This system is much less symmetric
and the wave functions for species 1 are more spread out due to the larger
intra-species interaction. Once again the Bloch state at the zone boundary,
$k=\kappa/2$, has nodes for both species. 
\begin{figure}
\includegraphics[width=8cm]{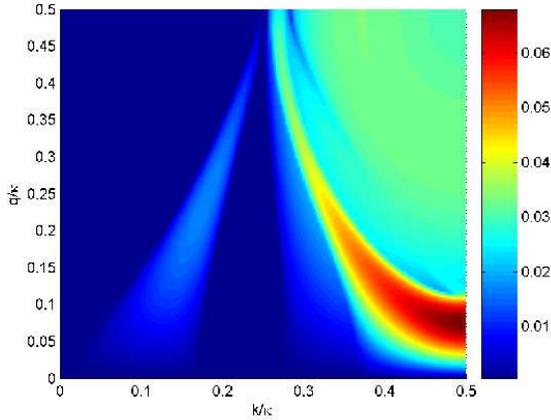}
\caption{Instability $\theta(q)$ of a Rb-Na condensate as a function of $k$ and $q$ for $v=0.6085E_R$ (Colour online).\label{Fig14}}
\end{figure}
Figure \ref{Fig14} is a colour map showing,
for each Bloch wavevector, $k$, and each fluctuation wavevector, $q$, the
instability $\theta(q)$. As can be seen, this map is qualitatively different to that of a
single component system. The map has two regions, small $k$ and larger $k$.
The large $k$ behaviour is similar to the one component case in that
instability sets in at $k\approx\kappa/4$ for modes with $q=\kappa/2$. At
larger $k$ the dominant (i.e. most rapidly growing) mode moves to longer
wavelengths. Unlike the single component case, the short wavelength modes do
not become stable as $k$ increases and at $k=\kappa/2$ all $q$'s are unstable.
The small $k$ regime is quite unlike the single component systems, there are
long wavelength unstable modes even at $k=0$ which persist up to a maximum
$k,$ with the dominant fluctuations moving towards shorter wavelengths. This
leaves a narrow window of Bloch wave vectors for which the system is stable.

The origin of the instability at small $k$ lies in the fact that this system
is unstable even in the absence of an optical lattice exhibiting a strong
tendency to phase separate into single component domains. As shown by \cite{Law, Goldstein} the
condition for the mode with wavevector $q$ to be stable in the absence of a
lattice is%
\begin{equation}
4q^{4}+2q^{2}n\left(  g_{11}+g_{22}\right)  +n^{2}\left(  g_{11}g_{22}%
-g_{12}^{2}\right)  >0
\end{equation}
so that if $g_{11}g_{22}<g_{12}^{2}$ only modes with sufficiently large $q$ are dynamically stable. It is clear that the parameters for the Na-Rb mixture do not satisfy this stability criterion.

\section{Using an optical lattice to stabilize two-component condensates}
The depth of the lattice potential used above was chosen rather arbitrarily. We expect that increasing the depth of the optical lattice should enhance the stability of the NaRb system at low $k$ because it confines the two components in different places. In particular, a very large value of $V_0$ should lead to an array of pure phase domains with the same periodicity as the optical lattice. In order to see how deep an optical lattice is required to stabilize the system we have carried out the stability analysis for the $k=0$ Bloch states for a range of values of $v$ for the NaRb mixture. 
\begin{figure}
\includegraphics[width=8cm]{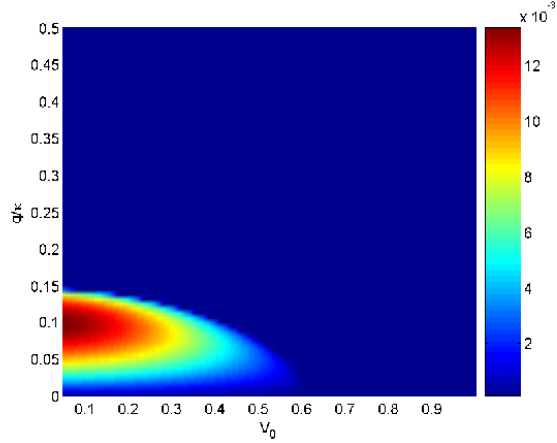}
\caption{A density map of the instability $\theta(q)$ for $k=0$ as a function of fluctuation wave-vector $q$ and optical lattice depth $V_0$ (colour online). \label{Fig13}}
\end{figure}
Figure \ref{Fig13} shows a contour plot of the instability $\theta(q)$ as a function of $V_0$ and the fluctuation wave-vector $q$. As can be seen the instability is indeed suppressed as $v$ increases. 
\begin{figure}
\includegraphics[width=8cm]{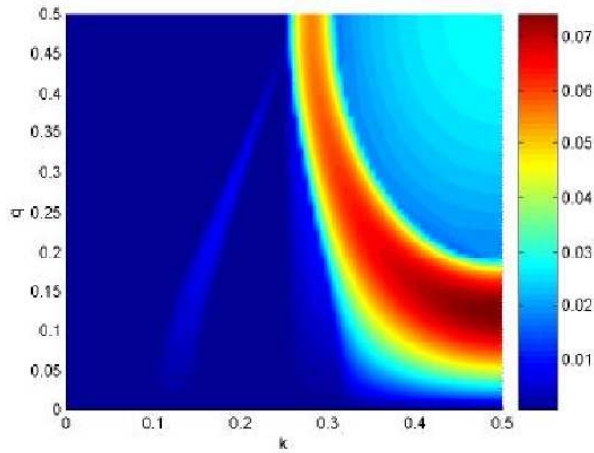}
\caption{Instability $\theta(q)$ of a Rb-Na condensate as a function of $k$ and $q$ for $V_0=1.217E_R$ (colour online). \label{Fig15}}
\end{figure}
In figure \ref{Fig15}  we show a contour map of the instability $\theta(q)$ as a function of $k$ and $q$ for $V_0=1.217E_R$ which shows that although the $k=0$ Bloch state has been stabilized, the instability re-appears at finite $k$ - a deeper optical lattice is required to stabilize a moving condensate. 
\begin{figure}
\includegraphics[width=8cm]{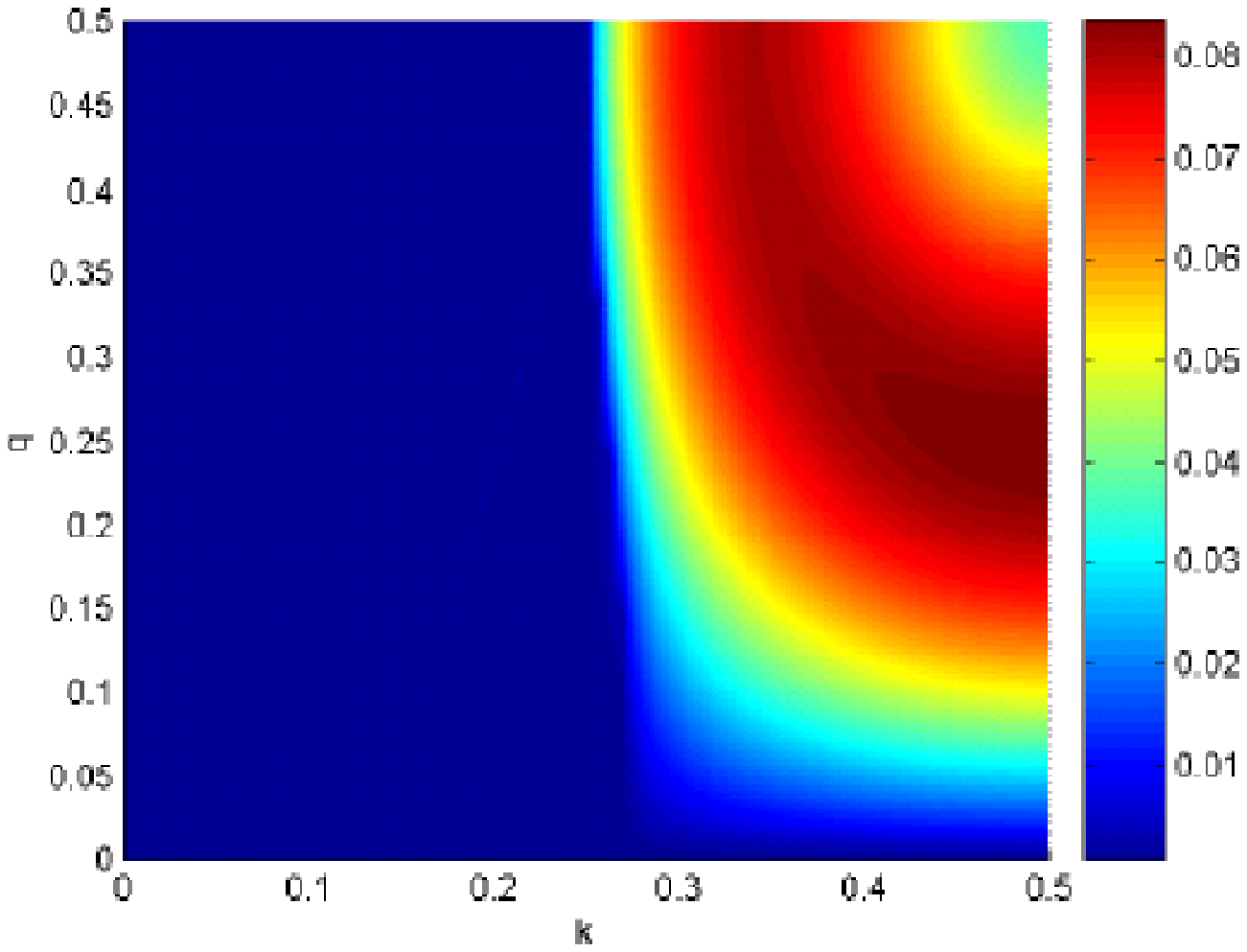}
\caption{Instability $\theta(q)$ of a Rb-Na condensate as a function of $k$ and $q$ for $V_0=2.434E_R$ (colour online).\label{Fig16}}
\end{figure}
Figure \ref{Fig16} shows the instability $\theta(q)$ as a function of $k$ and $q$ for $V_0=2.434E_R$. In this case the phase separation instability is fully suppressed. The price for this supression is that the modulational instability at large $k$ is much worse, with the dominant instability moved to higher $q$ but with all $q$ modes in the lowest two fluctuation bands being unstable once $k>\kappa/4$. An alternative way of showing the same physics is to plot the stability boundary for $k=0$ Bloch states as a function of $q$ and $ng_{11}$($=ng_{22}$) for fixed $ng_{12}=0.5E_R$ for a range of values of $v$ as shown in figure \ref{Fig12}.
\begin{figure}
\includegraphics[width=9cm]{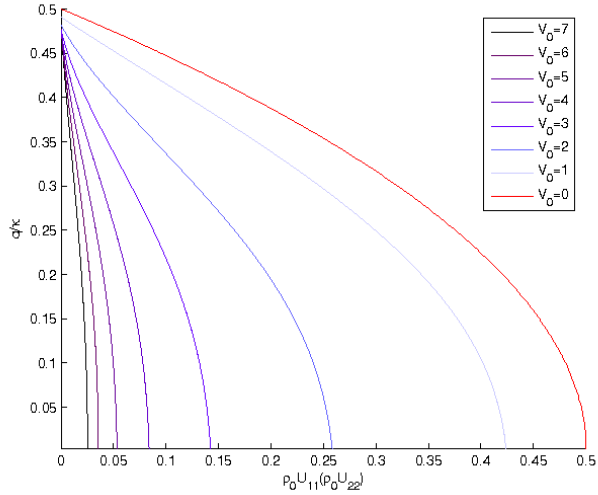}
\caption{The stability limit as a function for the largest unstable mode at $k=0$ for a condensate with fixed $ng_{12}=0.5E_R$ as a function of $ng_{11}=ng_{22}$ for values of $V_0$ between $0$ and $7E_R$ (colour online). \label{Fig12}}
\end{figure}
Increasing $v$ shrinks the region of instability below the line as one would expect.

Hence we have shown that two-component Bose condensates in 1d optical lattices can exhibit both the modulational instability at large Bloch wavevectors associated with single component systems and the instability to phase separation at small $k$ that can occur in the absence of a lattice. Turning on the lattice enhances the latter instability but suppresses the former. Hence such optical lattices could be used to hold immiscible two-component systems in intimate contact and to move them around within a trap.

\acknowledgments{SH thanks the EPSRC (UK) for financial support}

\bibliography{Coupled}

\begin{thebibliography}{9}
\expandafter\ifx\csname natexlab\endcsname\relax\def\natexlab#1{#1}\fi
\expandafter\ifx\csname bibnamefont\endcsname\relax
  \def\bibnamefont#1{#1}\fi
\expandafter\ifx\csname bibfnamefont\endcsname\relax
  \def\bibfnamefont#1{#1}\fi
\expandafter\ifx\csname citenamefont\endcsname\relax
  \def\citenamefont#1{#1}\fi
\expandafter\ifx\csname url\endcsname\relax
  \def\url#1{\texttt{#1}}\fi
\expandafter\ifx\csname urlprefix\endcsname\relax\def\urlprefix{URL }\fi
\providecommand{\bibinfo}[2]{#2}
\providecommand{\eprint}[2][]{\url{#2}}

\bibitem[{\citenamefont{Morsch and Oberthaler}(2006)}]{MorschReview}
\bibinfo{author}{\bibfnamefont{O.}~\bibnamefont{Morsch}} \bibnamefont{and}
  \bibinfo{author}{\bibfnamefont{M.}~\bibnamefont{Oberthaler}},
  \bibinfo{journal}{Rev. Mod. Phys.} \textbf{\bibinfo{volume}{78}},
  \bibinfo{pages}{179} (\bibinfo{year}{2006}).

\bibitem[{\citenamefont{Wu and Niu}(2003)}]{WuNiu}
\bibinfo{author}{\bibfnamefont{B.}~\bibnamefont{Wu}} \bibnamefont{and}
  \bibinfo{author}{\bibfnamefont{Q.}~\bibnamefont{Niu}}, \bibinfo{journal}{New
  J. Phys.} \textbf{\bibinfo{volume}{5}}, \bibinfo{pages}{104}
  (\bibinfo{year}{2003}).

\bibitem[{\citenamefont{Machholm et~al.}(2003)\citenamefont{Machholm, Pethick,
  and Smith}}]{Machholm}
\bibinfo{author}{\bibfnamefont{M.}~\bibnamefont{Machholm}},
  \bibinfo{author}{\bibfnamefont{C.}~\bibnamefont{Pethick}}, \bibnamefont{and}
  \bibinfo{author}{\bibfnamefont{H.}~\bibnamefont{Smith}},
  \bibinfo{journal}{Phys. Rev. A} \textbf{\bibinfo{volume}{67}},
  \bibinfo{pages}{053613} (\bibinfo{year}{2003}).

\bibitem[{\citenamefont{Pitaevskii and Stringari}(2003)}]{PitaevskiiBook}
\bibinfo{author}{\bibfnamefont{L.}~\bibnamefont{Pitaevskii}} \bibnamefont{and}
  \bibinfo{author}{\bibfnamefont{S.}~\bibnamefont{Stringari}},
  \emph{\bibinfo{title}{Bose-Einstein Condensation}}
  (\bibinfo{publisher}{Oxford University Press}, \bibinfo{year}{2003}).

\bibitem[{\citenamefont{Pitaveskii}(1961)}]{Pitaveskii61}
\bibinfo{author}{\bibfnamefont{L.}~\bibnamefont{Pitaveskii}},
  \bibinfo{journal}{Sov. Phys. JETP} \textbf{\bibinfo{volume}{13}},
  \bibinfo{pages}{451} (\bibinfo{year}{1961}).

\bibitem[{\citenamefont{Garay et~al.}(2001)\citenamefont{Garay, Anglin, Cirac,
  and Zoller}}]{Garay}
\bibinfo{author}{\bibfnamefont{L.}~\bibnamefont{Garay}},
  \bibinfo{author}{\bibfnamefont{J.}~\bibnamefont{Anglin}},
  \bibinfo{author}{\bibfnamefont{J.}~\bibnamefont{Cirac}}, \bibnamefont{and}
  \bibinfo{author}{\bibfnamefont{P.}~\bibnamefont{Zoller}},
  \bibinfo{journal}{Phys. Rev. A.} \textbf{\bibinfo{volume}{63}},
  \bibinfo{pages}{023611} (\bibinfo{year}{2001}).

\bibitem[{\citenamefont{Myatt et~al.}(1997)\citenamefont{Myatt, Burt, Ghrist,
  Cornell, and Wieman}}]{Myatt}
\bibinfo{author}{\bibfnamefont{C.}~\bibnamefont{Myatt}},
  \bibinfo{author}{\bibfnamefont{E.}~\bibnamefont{Burt}},
  \bibinfo{author}{\bibfnamefont{R.}~\bibnamefont{Ghrist}},
  \bibinfo{author}{\bibfnamefont{E.}~\bibnamefont{Cornell}}, \bibnamefont{and}
  \bibinfo{author}{\bibfnamefont{C.}~\bibnamefont{Wieman}},
  \bibinfo{journal}{Phys. Rev. Lett.} \textbf{\bibinfo{volume}{78}},
  \bibinfo{pages}{586} (\bibinfo{year}{1997}).

\bibitem[{\citenamefont{Law et~al.}(1997)\citenamefont{Law, Pu, Bigelow, and
  Eberly}}]{Law}
\bibinfo{author}{\bibfnamefont{C.}~\bibnamefont{Law}},
  \bibinfo{author}{\bibfnamefont{H.}~\bibnamefont{Pu}},
  \bibinfo{author}{\bibfnamefont{N.}~\bibnamefont{Bigelow}}, \bibnamefont{and}
  \bibinfo{author}{\bibfnamefont{J.}~\bibnamefont{Eberly}},
  \bibinfo{journal}{Phys. Rev. Lett.} \textbf{\bibinfo{volume}{79}},
  \bibinfo{pages}{3105} (\bibinfo{year}{1997}).

\bibitem[{\citenamefont{Goldstein and Meystre}(1997)}]{Goldstein}
\bibinfo{author}{\bibfnamefont{E.}~\bibnamefont{Goldstein}} \bibnamefont{and}
  \bibinfo{author}{\bibfnamefont{P.}~\bibnamefont{Meystre}},
  \bibinfo{journal}{Phys. Rev. A} \textbf{\bibinfo{volume}{55}},
  \bibinfo{pages}{2935} (\bibinfo{year}{1997}).

\end{thebibliography}

\end{document}